\documentstyle[12pt,aaspp4]{article}
\def\magdex{mag dex$^{-1}$}
\begin{document}

\title{A ``Short'' Distance to the Large Magellanic Cloud With the
{\em Hipparcos}\/ Calibrated Red Clump Stars}

\author{K. Z. Stanek\altaffilmark{1}}
\affil{Harvard-Smithsonian Center for Astrophysics, 60 Garden St., MS20,
Cambridge, MA 02138}
\affil{e-mail: kstanek@cfa.harvard.edu}
\altaffiltext{1}{On leave from Copernicus Astronomical Center, 
Warsaw, Poland}
\author{D. Zaritsky and J. Harris}
\affil{UCO/Lick Observatory and Department of Astronomy and Astrophysics,
University of California, Santa Cruz, CA 95064}
\affil{e-mail: dennis@ucolick.org, jharris@ucolick.org}

\begin{abstract}

Following an approach developed by Paczy\'nski \& Stanek, we derive a
distance to the Large Magellanic Cloud (LMC) by comparing red clump
stars from the {\em Hipparcos}\/ catalog with the red clump stars
observed in two fields in the LMC that were selected from the ongoing
photometric survey of the Magellanic Clouds to lie in low extinction
regions. The use of red clump stars allows a single step determination
of the distance modulus to the LMC, $\mu_{0,LMC} = 18.065\pm 0.031\pm
0.09\;$mag (statistical plus systematic error), and the corresponding
distance, $R_{LMC}= 41.02\pm 0.59\pm 1.74\;kpc$. This measurement is
in excellent agreement with the recent determination by Udalski et
al., also based on the red clump stars, but is $\sim 0.4\;$mag smaller
than the generally accepted value of $\mu_{0,LMC} = 18.50\pm
0.15\;$mag. We discuss possible reasons for this discrepancy and how
it can be resolved.

\end{abstract}


\section{INTRODUCTION}

The generally accepted distance modulus to the Large Magellanic Cloud
(LMC) is $\mu_{0,LMC} \approx 18.5 \pm0.15\;$mag (for recent
discussion see Westerlund 1997, Madore \& Freedman 1998).  However,
there is a long standing $\sim 0.3\;$mag discrepancy between the
``long'' distance determined using Cepheids (e.g. Laney \& Stobie
1994) and the ``short'' distance determined using RR Lyr stars
(e.g. Walker 1992, Layden et al.~1996). A similar discrepancy is present
in the distance to the LMC derived with the supernova SN1987A
($\mu_{0,LMC}< 18.37\;$mag, Gould \& Uza 1998; $\mu_{0,LMC} =
18.56\;$mag, Panagia et al.~1997).  Recently Udalski et al.~(1998)
used red clump stars observed in the LMC by the OGLE 2 project
(Udalski et al.~1997) and obtained a value of $\mu_{0,LMC} = 18.08\pm
0.03 \pm 0.12 \;$mag (statistical plus systematic error).  This
distance modulus is $\sim 0.4\;$mag smaller than the ``long'' distance
modulus used, for example, by the {\em HST}\/ Extragalactic Distance
Scale Key Project team (e.g.~Rawson et al.~1997 and references
therein).  Because errors in the distance to the LMC can propagate
into errors in such key quantities as distances, luminosities, masses,
and sizes of extragalactic objects, it is important to check the
result of Udalski et al.~(1998) using independent data, in order to
investigate possible systematic errors.

Red clump stars are the metal rich equivalent of the better known
horizontal branch stars, and theoretical models predict that their
absolute luminosity only weakly depends on their age and chemical
composition (Seidel, Demarque, \& Weinberg 1987; Castellani, Chieffi,
\& Straniero 1992; Jimenez, Flynn, \& Kotoneva 1998).  Indeed, the
absolute magnitude-color diagram from {\em Hipparcos}\/ data (Perryman
et al.~1997, their Figure~3) clearly shows a compact red clump -- the
variance in the $I$-band magnitude is only $\sim 0.15\;$mag (Stanek \&
Garnavich 1998; Udalski et al.~1998).

Despite their large number and the theoretical understanding of their
evolution, red clump stars have seldom been used as distance
indicators.  However, Stanek (1995) and Stanek et al.~(1994, 1997)
used these stars to map the Galactic bar.  Paczy\'nski \& Stanek
(1998) used the red clump stars observed by the OGLE project (Udalski
et al.~1993) to obtain the distance to the Galactic center.  Stanek \&
Garnavich (1998) used red clump stars observed by the {\em HST}\/ in
M31 to obtain a one-step distance to this galaxy.  In this paper we
follow the approach of Paczy\'nski \& Stanek (1998) and present an
estimate of the distance to the LMC based on the comparison between
the red clump giants observed locally by the {\em Hipparcos}\/
(Perryman et al.~1997) satellite and those observed in the LMC by the
$UBVI$ digital photometric survey of the Magellanic Clouds (Zaritsky
et al.~1997).  In Section 2 we describe the data used in this paper
and select low extinction regions for further analysis.  In Section 3
we analyze the red clump distribution in the LMC and derive the
distance to this galaxy. In Section 4 we discuss the possible reasons
for the discrepancy with the Cepheid distance to the LMC and how it
can be resolved.

\section{THE DATA}

\begin{figure}[t]
\plotfiddle{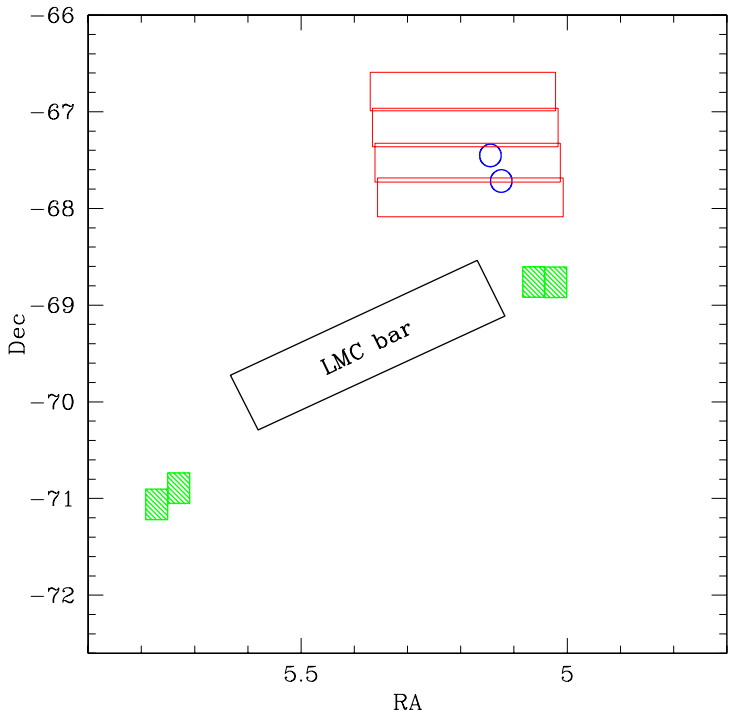}{8cm}{0}{50}{50}{-160}{-90}
\caption{Area of the LMC observed by Zaritsky et al.~(1997) (four
large parallel rectangles). Also shown are the two low-extinction
fields analyzed in this paper (small circles) and four fields used by
Udalski et al.~(1998) (shaded small rectangles) to determine the red
clump distance to the LMC. The position of the LMC bar is also roughly
marked.}
\label{fig:lmc}
\end{figure}

Zaritsky, Harris \& Thompson (1997) have undertaken a large scale
$UBVI$ photometric CCD survey of the Magellanic Clouds, with the
ultimate goal of imaging the central $8\arcdeg\times 8\arcdeg$ of the
LMC and $4\arcdeg\times 4\arcdeg$ of the SMC. The initial results for
a $\sim 2\arcdeg\times 1.5\arcdeg$ region were presented by Zaritsky
et al.~(1997), and the extinction map for the same region was
constructed by Harris, Zaritsky \& Thompson (1997). The area of the
LMC observed by Zaritsky et al.~(1997) is shown in
Figure~\ref{fig:lmc}, along with the four fields used by Udalski et
al.~(1998) to determine the red clump distance to the LMC.

As noted by Udalski et al.~(1998), uncertainties in the extinction
estimates are the largest contributor to their systematic error.  We
therefore select low reddening regions from the area observed by
Zaritsky et al.~(1997).  Harris et al.~(1997) used $\sim 2000$ OB main
sequence stars to construct a map of the reddening in the region
observed by Zaritsky et al.~(1997). They find a mean reddening of
$\langle E(B-V) \rangle_{LMC} = 0.20\;$mag, with a non-Gaussian tail
to high values.  As discussed by Harris et al.~(1997), it is possible
that the reddening map is biased to high reddening values, since it is
based on the reddening toward OB stars, a population that may reside
in dustier-than-average regions of the ISM.  On the other hand, the
nature of their interpolation may partially counter this bias by
smoothing over reddening spikes.

\begin{figure}[t]
\plotfiddle{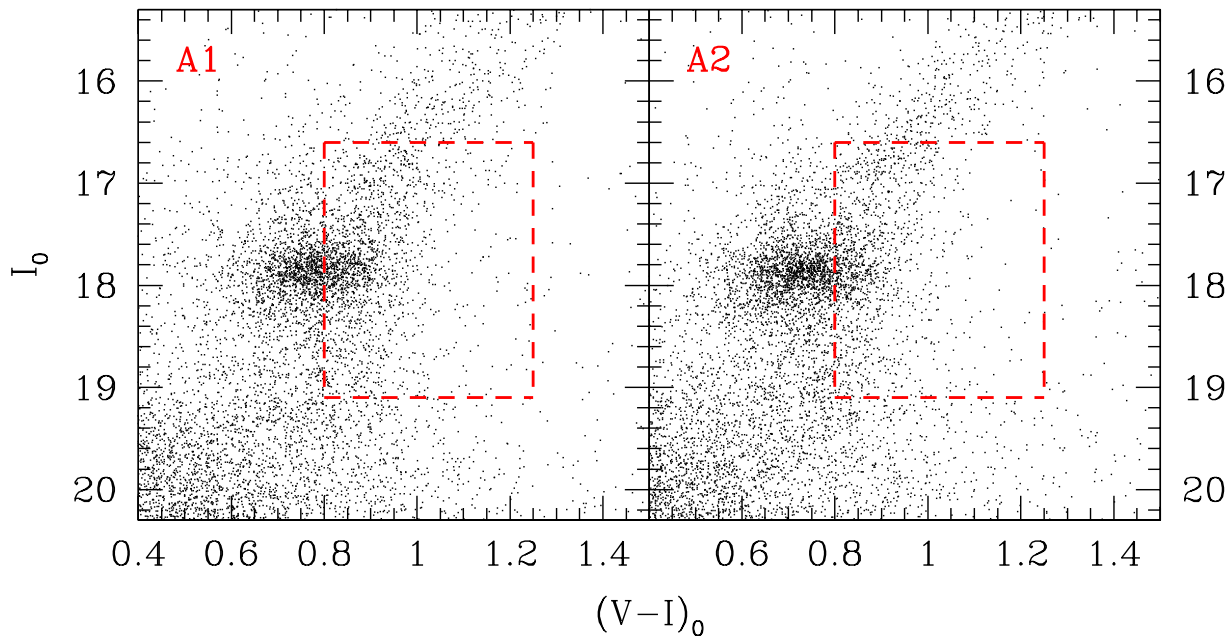}{7cm}{0}{80}{80}{-250}{-320}
\caption{The red clump dominated parts of CMDs for the two fields A1
and A2 in the LMC, selected from the survey of Zaritsky et al.~(1997).
The dashed rectangles surround the red clump region used for the
comparison between the local (observed by {\em Hipparcos}\/ and the
LMC red clump stars.}
\label{fig:cmd}
\end{figure}

Recently Schlegel, Finkbeiner \& Davis (1998; hereafter: SFD)
published a new all-sky reddening map, based on the COBE/DIRBE and
IRAS/ISSA maps. With its high spatial accuracy, the SFD map might be
potentially useful for selecting low-reddening regions in the
Magellanic Clouds. However, as discussed by SFD, the LMC, SMC and M31
are not removed from their map, nor are sources within their Holmberg
radius.  The reddening estimates from the SFD map are unreliable
within these objects because of the lack of spatial temperature
resolution from DIRBE and confusion with internal IR sources. However,
because neither of these two effects artificially creates large areas
of low redenning in the map, the map can be used as a guide to select
regions of low total extinction. If the total column density of dust
is low, then the dispersion of reddening values of a population
distributed along the line-of-sight will also be low. We therefore
decided to select regions of low DIRBE/IRAS reddening from the area
observed by Zaritsky et al.~(1997) and then determine the reddening
using the map of Harris et al.~(1997).

Examining the SFD map in the region observed by Zaritsky et al.~(1997)
we selected two circular regions with $7\arcmin$ radii (defined by the
low extinction region), which we hereafter call A1 and A2. These
regions are centered on (RA,Dec)$=(5.1442^h,-67.452^\circ)$ and
(RA,Dec)$=(5.1239^h, -67.716^\circ)$, respectively, and are shown as small
circles in Figure~\ref{fig:lmc}.  The SFD map gives an average
reddening of $E(B-V) \approx 0.18\;$mag for each of these regions. The
Harris et al.~(1997) map gives $E(B-V)=0.16$ for the region A1 and
$E(B-V)=0.17$ for the region A2. We find the good agreement between
the maps reassuring.  However, we assume a conservative error of the
$E(B-V)$ reddening to be $0.04\;$mag, following the discussion by
Harris et al.~(1997).  This leads to the values of the extinction
$A_I=0.31\pm 0.08\;$mag and reddening $E(V-I)=0.20\pm 0.05\;$mag for
the region A1 and $A_I=0.33\pm 0.08\;$mag and $E(V-I)=0.22\pm
0.05\;$mag for the region A2.

In Figure~\ref{fig:cmd} we show the red clump dominated parts of the
$I_0,\;(V-I)_0$ color-magnitude diagrams (CMDs) for both regions A1
and A2, corrected for the extinction and the reddening using the above
values of $A_I$ and $E(V-I)$. The dashed rectangle corresponds to the
region of the CMD selected for comparison with the local red clump
stars observed by {\em Hipparcos}\/ (see the next Section).  The LMC
red clump is clearly bluer than the local one (Paczy\'nski \& Stanek
1998, their Figure~2), indicating a lower average metallicity of the
LMC (see Jimenez et al. 1998, their Figure~5). However, there is a
sufficient overlap with the {\em Hipparcos}\/ color range to allow for
meaningful comparison, which we perform in the next Section.

\section{THE ANALYSIS}

Following Paczy\'nski \& Stanek (1998), we selected the red clump
stars in the color range $0.8<(V-I)_0<1.25$ and the magnitude range
$16.6<I_0<19.1$ in the A1 region (1725 stars) and in the A2 region
(1273 stars).  The color range was selected to correspond to the color
range of the red clump stars observed locally by the {\em Hipparcos}
(Paczy\'nski \& Stanek 1998, their Figure~2). Following Garnavich \&
Stanek (1998), we fitted both distributions with a function
\begin{equation}
n(I_0) = a + b (I_0-I_{0,m})  + c (I_0-I_{0,m})^2 +
\frac{N_{RC}}{\sigma_{RC}\sqrt{2\pi}}
 \exp\left[-\frac{(I_0-I_{0,m})^2}{2\sigma_{RC}^2} \;\right].
\end{equation}
The first three terms describe a fit to the ``background''
distribution of the red giant stars, and the Gaussian term represents
a fit to the red clump itself.  $I_{0,m}$ corresponds to the peak
magnitude of the red clump population. We obtained the values of
$I_{0,m}=17.832 \pm0.012$ for the A1 region and $I_{0,m}=17.843
\pm0.020$ for the A2 region.

\begin{figure}[t]
\plotfiddle{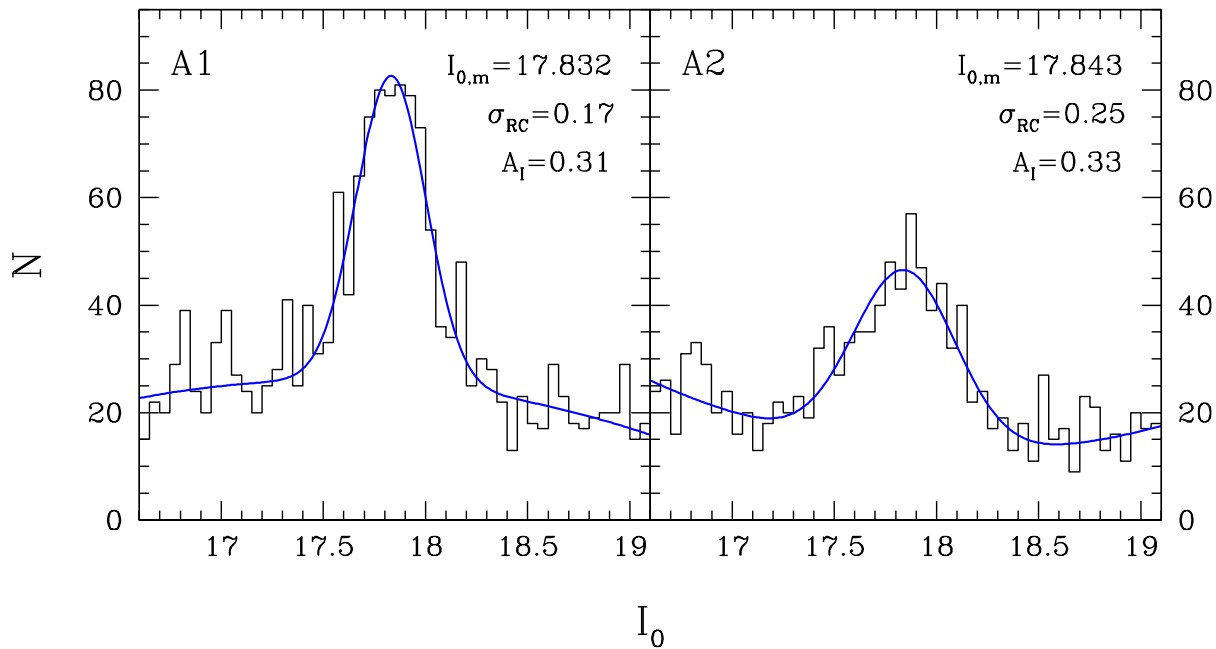}{7cm}{0}{80}{80}{-250}{-320}
\caption{The distribution of the LMC red clump stars in the regions A1
and A2 as a function of their $I_0$ magnitude, along with the
analytical fits described by the Eq.1.}
\label{fig:dist}
\end{figure}

The distribution of the LMC red clump stars as a function of their
$I_0$ magnitude is shown in Figure~\ref{fig:dist} along with the
fitting function described by the Eq.1.  The Gaussian fitted to the A1
field red clump distribution has a smaller dispersion, $\sigma_{RC}=
0.17\;$mag, than the Gaussian fitted to the A2 red clump,
$\sigma_{RC}=0.25\;$mag.  The red clump is less pronounced, and has a
correspondingly larger dispersion, in the A2 field because the red
clump is bluer and mostly falls outside the color cut.  Indeed, when
we select the red clump stars within the color range
$0.55<(V-I)_0<0.8$, both fitted distributions have the same
$\sigma_{RC}= 0.18\;$mag. This different color selection also moves
the red clump peaks $I_{0,m}$ to slightly fainter magnitudes,
$I_{0,m}=17.876 \pm0.008$ for the A1 region (1907 stars) and
$I_{0,m}=17.891 \pm0.007$ (2497 stars) for the A2 region.  This slight
color dependence in the $I$-band magnitudes of the red clump stars is
consistent with the results of Paczy\'nski \& Stanek (1998), Stanek \&
Garnavich (1998) \& Udalski et al.~(1998).

We now proceed to obtain the LMC distance modulus using the red clump,
by assuming that the absolute $I$-band brightness of the red clump
stars is the same for the local stars observed by {\em Hipparcos}\/
and those in the LMC. The straight average of the red clump peak
apparent magnitudes $I_{0,m}$ for the two regions is $\langle I_{0,m}
\rangle = 17.837 \pm 0.008$ and the weighted mean is $\bar{I}_{0,m} =
17.835 \pm 0.008$.  Combining $\bar{I}_{0,m}$ with the distribution of
local red clump stars, which have $M_{I,m} = -0.23\pm0.03$ (Stanek \&
Garnavich 1998), we obtain the distance modulus for the LMC,
$\mu_{0,LMC} = 18.065 \pm 0.031\;$mag, or $R_{LMC}=41.02 \pm
0.59\;kpc$ (statistical error only).  After adding the systematic
error of $0.08\;$mag due to the uncertainty in the $A_I$
determination, and $0.04\;$mag due to the zero-point uncertainty in
the $I$-band photometry (Zaritsky et al.~1997), we arrive at the final
value of $\mu_{0,LMC}= 18.065 \pm 0.031 \pm 0.09\;$mag (statistical
plus systematic error). This is indistinguishable from the value of
$\mu_{0,LMC}= 18.08 \pm 0.03 \pm 0.12\;$mag determined by Udalski et
al.~(1998), but $\sim 0.4\;$mag below generally accepted value of
$\mu_{0,LMC}= 18.50 \pm 0.15\;$mag (Madore \& Freedman 1998). We
discuss possible reasons for this discrepancy and how it can be
resolved in the next Section.

\section{DISCUSSION}

As with all distance-ladder techniques, our analysis includes the
assumption that the calibrating and target objects being compared are
intrinsically similar. In our red clump analysis, this assumption is
manifested by the assertion that the $I$-band brightness of red clump
stars is independent of the age, chemical composition, and mass
differences that may exist between the red clump stars near the Sun
and those in the LMC.  Indeed, the LMC red clump is systematically
bluer than the local one, indicating the somewhat different properties
of these stars. However, Paczy\'nski \& Stanek (1998), Stanek \&
Garnavich (1998) and Udalski et al.~(1998) found that the $I$-band
peak magnitude of the red clump depends very weakly on their $(V-I)_0$
color in the range $0.7<(V-I)_0<1.4$, and therefore is independent of
the metallicity (Jimenez et al.~1998).  This is confirmed in this
paper as well, by comparing the peak brightness of the red clump for
two color ranges $0.55<(V-I)_0 <0.8$ and $0.8<(V-I)_0 <1.25$ (see the
previous Section). The fact that the observed red clump distributions
are so narrow ($\sigma_{RC}\approx 0.15\;$mag) indicates that the age
dependence of the red clump $I$-band peak luminosity is also small
($\lesssim 0.1\;$mag).  Otherwise, in a system with a complex star
formation history, such as the LMC (Holtzman et al.~1997; Geha et
al.~1998), the resulting red clump should have considerable width.
Stanek \& Garnavich (1998) compared three different lines-of-sight
that probe a large range of M31 galactocentric distances and
locations, and hence a range of metallicities and possibly ages and
star formation histories. The fact that the derived distance moduli
for their three fields varied by only $\sim 0.035\;$mag indicates that
the red clump is a potentially stable standard candle. The mostly
empirical support for using the red clump stars as a distance
indicator should also be verified using modern theory of the stellar
structure and evolution. In particular, $I$-band predictions are
seldom given by such theoretical calculations.

So why does the red clump distance to the LMC disagree with the
Cepheid distance (Madore \& Freedman 1998)?  As usual, there are
several possible answers. Contrary to our arguments given above, there
might still be something ``unusual'' about the red clump population in
the LMC.  Although the red clump and Cepheid distances to the LMC are
discrepant, the distances to M31 derived from the two methods are in
excellent agreement ($m-M = 24.471\pm0.035\pm0.045$ from Stanek \&
Garnavich~1998 and 24.44$\pm 0.13$ from Freedman \& Madore~1990).
Another possibility is that the Cepheid distance to the LMC is simply
poorly determined, as there are few Cepheids with well determined
parallaxes in the {\em Hipparcos}\/ catalog.  In their recent study,
Madore \& Freedman (1998) find $\mu_{0,LMC}= 18.44\pm 0.35\;$mag, from
a sample of 19 Cepheids observed by {\em Hipparcos}\/ with good $BV$
data, and $\mu_{0,LMC}= 18.57\pm 0.11\;$mag, from a sample of only 7
Cepheids with good $BVIJHK$ data. Yet a third possibility, as
discussed by Madore \& Freedman (1998), is that there are other
effects on the Cepheid PL relation (e.g. extinction, metallicity, and
statistical errors), which are as significant as any reassessment of
the zero point based on {\em Hipparcos}. The metallicity effect on the
Cepheid PL relation, determined by Kennicutt et al.~(1998) ($\delta(m
- M)_0/\delta[O/H] = -0.24 \pm 0.16 $\magdex), reduces the discrepancy
between the red clump and Cepheid distances to the LMC by $\sim
0.1\;$mag, while the somewhat larger metallicity dependence found by
Sasselov et al.~(1997) and Kochanek (1997) reduces it by $\sim
0.15\;$mag.  To illustrate the effect of the assumed reddening on the
derived distance modulus, we note that the value of the LMC distance
modulus, $\mu_{0,LMC}= 18.54\;$mag, derived recently by Salaris \&
Cassisi (1998) and based on the $V$-band brightness of the RR Lyr
stars, becomes $\mu_{0,LMC}= 18.22\;$mag if their assumed reddening of
$E(B-V)=0.10\;$mag is increased to $E(B-V)=0.20\;$mag, corresponding
to the mean reddening found by Harris et al.~(1997). It is disturbing
that the distance to a key calibrator of the entire distance scale is
uncertain by up to 20\%.

As described by Udalski et al.~(1998), the $\sim 0.4\;$mag discrepancy
between their (and now our as well) ``short'' distance to the LMC and
the ``long'' distance to the LMC from the Cepheids can be resolved by
using detached eclipsing binaries as a direct distance indicator
(Paczy\'nski 1997). The Cepheids in the LMC can also be used to get a
direct distance estimate through a modified Baade-Wesselink method
(e.g. Krockenberger 1996; Krockenberger, Sasselov, \& Noyes 1997).
Both these methods require no intermediate steps in the distance
ladder, therefore avoiding the propagation of errors usually crippling
the distance scale.  With the 6.5--8 meter telescopes now being built
in the Southern Hemisphere the necessary spectroscopy of the detached
eclipsing binaries and Cepheids can be quite easily obtained for these
14--18 magnitude stars. It is worth mentioning here that the effort to
obtain direct distances with the detached eclipsing binaries and
Cepheids to the M31 and M33 galaxies is already under way and the
first results look promising (project DIRECT: Kaluzny et al.~1998,
Stanek et al.~1998, Krockenberger et al.~1998, Sasselov et al.~1998).

To summarize, among the various stellar distance indicators the red
clump giants might be the best for determining the distance to the LMC
and other nearby galaxies because there are so many red clump stars.
In particular, {\em Hipparcos}\/ provided accurate distance
determinations for almost 2,000 such stars, but unfortunately $I$-band
photometry is available for only $\sim 30\%$ of them, so it is
important to obtain $I$-band photometry for all {\em Hipparcos}\/ red
clump giants.  We also need to test the metallicity dependence of {\em
Hipparcos}\/ red clump giant absolute luminosities and colors. There
are many stars within $100\;pc$ of the Sun for which very
high-resolution spectroscopy is possible.

\acknowledgments{KZS was supported by the Harvard-Smithsonian Center
for Astrophysics Fellowship. DZ acknowledges financial support for the
Magellanic Cloud Photometric Survey and related science from a NASA
LTSA grant (NAG-5-3501), an NSF grant (AST-9619567), and a David
and Lucille Packard Foundation Fellowship.}

\end{document}